\documentclass[aps,rmp,reprint,showkeys]{revtex4-1}
\usepackage{amsmath}
\usepackage{amsfonts}
\usepackage{amssymb}
\usepackage{color}
\usepackage{graphicx}
\usepackage[version=4]{mhchem}
\usepackage{chemfig}
\newtheorem{theorem}{Theorem}
\usepackage{subcaption}
\begin{document}

\title{Long-term behaviours of Autocatalytic sets}
\author{Alessandro Ravoni}
\affiliation{Department of Mathematics and Physics, University of Roma Tre, Via della Vasca Navale 84, 00146 Rome, Italy}

\begin{abstract}
Autocatalytic Sets are reaction networks theorised as networks at the basis of life.
Their main feature is the ability of spontaneously emerging and self-reproducing.
The Reflexively and Food-generated theory provides a formal definition of Autocatalytic Sets in terms of graphs with peculiar topological properties.
This formalisation has been proved to be a powerful tool for the study of the chemical networks underlying life, and it was able to identify autocatalytic structures in real metabolic networks.
However, the dynamical behaviour of such networks has not been yet complitely clarified.
In this work, we present a first attempt to connect the topology of an Autocatalytic Set with its dynamics.
For this purpose, we represent Autocatalytic Sets in terms of Chemical Reaction Networks, and we use the Chemical Reaction Network theory to detect motifs in the networks'structure, that allow us to determine the long-term behaviour of the system.
\end{abstract}



\maketitle

\section{Introduction}

\textit{Autocatalytic Sets} (ASs) are sets of chemical species that mutually catalyse each other's production through chemical reactions, starting from a basic food source.
Introduced by Kauffman \cite{Kauffman1971,Kauffman1986,Kauffman1993}, the notion of AS captures the idea of a system capable of spontaneously emerging and self-reproducing, using the resources provided by the environment.
These features are commonly accepted as basic qualities of early life forms \cite{Luisi2016,Kauffman1993,Higgs2015,Nghe2015,Rasmussen2016}, and many authors have suggested that the appearance of ASs can have been a fundamental step in the transition from non-living to living matter, possibly preceding RNA-based systems \cite{Kauffman2011,Hordijk2010,Hordijk2018a,Dyson1985,Vasas2012,Xavier2020}.
However, some properties of ASs are still under debate, such as their effective ability to form spontaneously \cite{Lifson1997,Szathmary2000} and undergo adaptive evolution \cite{Vasas2010,Vasas2012}, a feature that is generally referred to as \textit{evolvability}.

A formal definition of ASs has been proposed in the context of graph theory (see Section \ref{RAF_def} for details) by the \textit{Reflexively Autocatalytic and Food-generated} (RAF) theory \cite{Hordijk2004,Hordijk2011}. This allowed to overcome some limits associated with ASs and to make evident their relevance in the context of the origin of life.
In particular, RAF theory has successfully proved that ASs (or \textit{RAF sets}, using the nomenclature of RAF theory) are highly likely to exist in a catalytic reaction system \cite{Hordijk2004, Mossel2005,Hordijk2011, Hordijk2012,Vasas2012}, and RAF sets have been detected in the metabolic networks of Escherichia Coli \cite{Sousa2015} and ancient anaerobic autotrophs \cite{Xavier2020}. 
Moreover, RAF theory shows that RAF sets are often composed by RAF subsets, \cite{Hordijk2018a, Hordijk2012b,Hordijk2014}, and it has been argued that (some of) these subsets could be the elementary units on which natural selection may act \cite{Hordijk2014, Hordijk2012b,Vasas2012,Hordijk2018a,Hordijk2018b}.
First numerical results support this scenario, at least for compartmentalised RAF sets \cite{Vasas2012,Hordijk2018b, Serra2019,Ravoni2020}.
However, in the same works it has been observed that the conditions a graph must satisfy to be a RAF set (see Section \ref{RAF_def}) are not sufficient to strictly constrain its dynamics, and the system may not exhibit the features necessary to be evolvable. More generally, the connection between the topology of a RAF set and its dynamics is not yet fully understood \cite{Filisetti2011,Serra2019,Hordijk2018c,Hordijk2012a}, 

In this work we make a first attempt to solve this issue, by studying RAF sets through the formalism of the \textit{Chemical Reaction Network} (CRN) theory \cite{Feinberg2019}.
The CRN theory offers powerful tools that allow  to connect the topology of a network with its dynamical behaviour, for a large class of network dynamics \cite{Feinberg2019,Feinberg1987,Feinberg1995,Ellison1998,Ji2011}.
Note that the CRN theory usually deals with the study of real-world chemical systems.
Starting from the results of the CRN theory, we identify which conditions on the structure of RAF sets are useful for predicting their long-term behaviours.
In particular, we investigate which conditions determine a dynamics with desired properties for systems at the origin of life (e.g., self-maintenance, concentrations growth, homeostasis), adding new elements in the scenario previously proposed for the evolvability of RAF sets \cite{Hordijk2018b,Vasas2012}.

The paper is organised as follows. In Section \ref{Background} we recall notions of the CRN theory useful for our study, and the definition of RAF sets. In Section \ref{CRN_RAF} we show our representation of RAF sets in terms of CNRs. In Section \ref{motif} we introduce motifs in the structure of RAF sets that are connected with long-term behaviours of the networks, and we provide an illustrative example of evolvability of RAF sets by numerically simulate the dynamics. Finally, in Section \ref{conclusions} we discuss the conclusions.

\section{\label{Background}Background}
\subsection{Chemical Reaction Networks and the Deficiency Zero Theorem}

In this Section we recall some useful definitions and introduce the \textit{Deficiency Zero Theorem} (DZT) \cite{Feinberg1987,Feinberg1995}.
An interested reader can find a detailed discussion of the CRN theory in \citet{Feinberg2019}.
Let $\mathcal{S}$  
denote the set of chemical \textit{species},
$\mathcal{N} = \{\nu_i, \nu_i': i = 1,\cdots, r,\,\nu_j \neq \nu_i'\}$ the set of \textit{complexes}, that is, the members of the vector space $\mathbb{R}_{>0}^S$ generated by the species that provide the inputs and the outputs of a reaction, and $\mathcal{R} = \{\nu_i \rightarrow \nu_i': i = 1,\cdots, r\}$ the set of \textit{reactions}, that is, the set of relations among complexes. 
A CRN is a triple $(\mathcal{S}, \mathcal{N}, \mathcal{R})$.
Note that it is possible to consider \textit{open} CRNs introducing pseudo reactions such as $0 \rightarrow \nu$ and $\nu \rightarrow 0$, where $0$ is called the \textit{zero complex} and it is the zero vector of $\mathbb{R}^\mathcal{S}$.

We denote with $\mathcal{G}$ the directed graph with complexes as nodes and reactions $\nu_i \rightarrow \nu_i'$ as directed edges.
The connected components of $\mathcal{G}$ are called the \textit{linkage classes} of the CRN.
A \textit{strong-linkage class} is a strongly connected linkage class.
A \textit{terminal strong-linkage class} is a strong-linkage class containing no complex that is a source for an edge pointing to a different strong-linkage class.
Note that a complex not involved in any reaction is a terminal strong-linkage class.
A CRN is said to be \textit{weakly reversible} if each linkage class in $\mathcal{G}$ is a strong-linkage class, or, equivalently, if each linkage class is a terminal strong-linkage class.
The \textit{reaction vectors} for a CRN are the members of the set $\{ \nu_i' - \nu_i \in \mathbb{R}^\mathcal{S} | \nu_i \rightarrow \nu_i' \in \mathcal{R}\}$.
The span generated by the reaction vectors is called the \textit{stoichiometric subspace} of the network, and its dimension is denote by $d$.
The \textit{deficiency} of a CRN is:
\begin{equation}
 \delta = n - l - d,
\end{equation}
where $n$ and $l$ are the number of complexes and linkage classes, respectively.
Note that the dimension $d$ of the span of two networks with the same complexes and the same linkage classes is the same.
This implies that the two networks have the same deficiency $\delta$, and that it results $\delta \geq 0$ \cite{Feinberg1995}.

In this work we refer to \textit{mass-action systems}, that are CRNs taken together with an element $\kappa \in \mathbb{R}_{>0}^{\mathcal{R}}$.
The number $\kappa_{\nu \rightarrow \nu'}$ is the rate constant for the reaction $\nu_i \rightarrow \nu'_i \in \mathcal{R}$.
Hereafter, we use $\mathcal{G}$ to indicate both a CRN, its associated directed graph and the resulting mass-action system.
The \textit{species-formation rate function} for a mass-action system is the function $f(\cdot,\kappa): \mathbb{R}_{\geq 0 }^{\mathcal{S}} \rightarrow \mathbb{R}^\mathcal{S}$ defined by:
\begin{equation}
 f(c,\kappa) = \sum_{\nu \rightarrow \nu'} \kappa_{\nu \rightarrow \nu'} c^{\nu} (\nu'-\nu),
\end{equation}
where $c$ is the (time-dependent) vector of species concentration and $c^{\nu}$ is defined as:  
\begin{equation}
 c^{\nu} := \prod_{s \in \mathcal{S}} c(s)^{\nu(s)}.
\end{equation}
Here $c(s)$ is the concentration of species $s$ and $\nu(s)$ is the $s$-th component of the vector $\nu$, that is, the \textit{stoichiometric coefficient} of species $s$ in the complex $\nu$.
In fact, for each reaction $\nu_i \rightarrow \nu'_i \in \mathcal{R}$, the component $\nu_i(s)$ indicates the number of molecules of the $s$-th chemical species consumed by the $i$-th reaction, while the component $\nu'_i(s)$ indicates the number of molecules of the $s$-th species produced by the $i$-th reaction.
The dynamics of $\mathcal{G}$ is described by the equation:
\begin{equation}\label{ODE}
 f(c,\kappa) = \dot{c},
\end{equation}
where the point indicates the time derivative.
Eq.~\ref{ODE} correspond to a set of ordinary differential equations (ODEs) that describe the time evolution of concentration of each species in $S$.
Indeed, it is natural to define a \textit{positive equilibrium} as an element $c \in \mathbb{R}_{>0}^{\mathcal{S}}$ such that $f(c,\kappa) = 0$.
A positive equilibrium is then a steady state of the mass-action system characterized by a positive concentration and a zero net production rate for every species of the CRN.
Note that a necessary condition for the existence of a positive equilibrium for a mass-action system is the existence of a \textit{positive vector} $\alpha \in \mathbb{R}_{>0}^\mathcal{R}$ such that \cite{Feinberg1995}:
\begin{equation}\label{positive}
\sum_{\nu \rightarrow \nu'} \alpha_{\nu \rightarrow \nu'} (\nu'-\nu) = 0.
\end{equation}
If exists $\alpha$ such that Eq.~\ref{positive} is satisfied, we say that the network's reaction vectors are positively dependent.
Note that weak reversibility is a sufficient (but not necessary) condition for the existence of a positive equilibrium \cite{Boros2019,Feinberg1995}.

The DZT provided a connection between the topology of a network, expressed in terms of its deficiency, and the existence and the uniqueness of steady states for the associated mass-action system \cite{Feinberg1987,Feinberg1995}.
In fact, the DZT does much more, and its results can be applied to a broad class of kinetics systems.
However, in this work we limit our attention on a specific result (again, see \citet{Feinberg2019} for a comprehensive illustration of the theorem and its application).
In particular, we will use the following statement:
\begin{theorem}[The deficiency zero theorem]\label{zero_deficiency}
Let $(\mathcal{S},\mathcal{N},\mathcal{R})$ be a reaction network of deficiency zero.
\begin{enumerate}
 \item If the network is weakly reversible, the resulting mass-action system admits precisely one positive equilibrium for each initial condition, no matter what (positive) rate constants are assigned to the reactions~\label{i}.
 \item If the network is not weakly reversible, there can be no assignment of (positive) rate constants such that the resulting mass-action system admits the existence of a positive equilibrium (the network's reaction vectors are not positively dependent)\label{ii}.
 \end{enumerate}
\end{theorem}

\subsection{\label{RAF_def}Reflexively Autocatalytic and Food-generated Sets}

Within the framework of RAF theory \cite{Hordijk2004,Hordijk2011}, a network of interacting species is represented by a \textit{Catalytic Reaction Systems} (CRS) \cite{Lohn1998,Hordijk2004,Hordijk2017}, that is a tuple  $(S, R, C, F)$ such that:
\renewcommand{\labelitemi}{\textendash}
\begin{itemize}
 \item $S$ is the set of chemical species;
 \item $R$ is the set of reactions, $\rho \rightarrow \pi$, where $\rho, \pi \subset S$ are the \textit{reactants} and \textit{products} of a reaction, respectively;
 \item $C$ is the \textit{catalysis} set, that is, a set of pairs $\{(s,r):s\in~S, r\in R\}$ indicating the species $s$ as the \textit{catalyst} of reaction $r$;
 \item $F \subset S$ is the \textit{food} set, that is a distinguished subset of $S$ such that each species $s \in F$ is assumed to be available from the environment.
\end{itemize}

Let $R'$ be a subset of $R$. The \textit{closure} $cl_{R'}(F)$ is defined to be the (unique) minimal subset of $S$ that contains $F$ together with all species that can be produced from $F$ by repeated applications of reactions in $R'$. Note that $cl_{R'}(F)$ is well defined and finite \cite{Hordijk2004}. 
In RAF theory, the notion of ASs in represented by the RAF sets.
Given a CRS $(S,R,C,F)$, a RAF set is a set of reactions $R'\subseteq R$  (and associated chemical species) that satisfies the following properties:
\renewcommand{\labelitemi}{\textendash}
\begin{itemize}
\item Reflexively autocatalytic: for each reaction $r \in R'$ there exists at least one species $s \in cl_{R'}(F)$ such that $(s,r) \in C$; 
\item $F$-generated: for each reaction $r: \rho \rightarrow \pi, r \in R'$ and for each species $s \in S$ such that $s \in \rho$, it is $s \in cl_{R'}(F)$. 
\end{itemize}
Thus, a RAF set is a set of reactions able to catalytically produce all its reactants starting from a suitable food set.
It is also possible to define the closure of a set of reactions, introducing the notion of \textit{closed RAF set} \cite{Smith2014}.
Given a CRS $(S,R,C,F)$, a subset $R' $ of $R$ is said to be a closed RAF set if: 
\renewcommand{\labelitemi}{\textendash}
\begin{itemize}
\item $R'$ is a RAF set; 
\item for each $r$ such that all its reactants and at least one catalyst are either part of the set $F$ or are produced by a reaction from $R'$, it is $r \in R'$.
\end{itemize}
Fig.~\ref{FIG:1} shows an example of a RAF set and its constituent closed RAF sets.

In \citet{Smith2014}, authors claimed that ``informally, a closed RAF captures the idea that any reaction that can occur, will occur''.
It has been shown that closed RAF sets have interesting dynamical properties and they may be the relevant units for the potential evolvability of ASs \cite{Hordijk2018b}.
However, the structure of a closed RAF set seems unable to guarantee its dynamical stability, and monotonic growth of all its associated species \cite{Serra2019,Ravoni2020,Dittrich2007}.
In Section~\ref{motif} we introduce further constraints on the topology of closed RAF sets that allow us to predict their dynamics and to identify networks with peculiar dynamical properties.
\begin{figure}
\centering
\schemestart
 \chemfig{\textit{f1+f2}}
    \arrow(f1f2--x1){->[\textcolor{blue}{{\textit{(x2)}}}]}[0]
    \chemfig{\textit{x1}}
    \arrow{->[*{0}\textcolor{blue}{\textit{(x4)}}]}[270]
    \chemfig{\textit{x2+x3}}
    \arrow{<=>[\textcolor{blue}{\textit{(x1)}}]}[180]
    \chemfig{\textit{x4}}
    \arrow(x4--x1){->[*{0.0}\textcolor{blue}{\textit{(x3)}}]}[40,1.6]
    \arrow(@x4--xy){<=>[*{0}\textcolor{blue}{\textit{(x4)}}]}[270]
    \chemfig{\textit{xy}}
    \arrow(@f1f2--y1){->[\textcolor{blue}{\textit{(y3)}}]}[170]
    \chemfig{\textit{y1}}
    \arrow(@f1f2--y2){->[][\textcolor{blue}{\textit{(y1)}}]}[190]
    \chemfig{\textit{y2}}
    \arrow(@xy--y3){<-[\textcolor{blue}{\textit{(x1)}}]}[180]
    \chemfig{y3}
    \arrow{<=>[][*{0}\textcolor{blue}{\textit{(y2)}}]}[90]
    \chemfig{\textit{y1+y2}}
\schemestop
\caption{\label{FIG:1}RAF set example. Complexes are represented by black letters (the letter $f$ indicates food species), reactions by arrows.
Blue letters in brackets indicate catalysts.
The entire reaction network is a RAF set.
Two closed RAF sets are present within the RAF set, namely: 
$R^{(1)} = \{\ce{\textit{f1+f2} ->[\textcolor{blue}{\textit{(x2)}}] \textit{x1}},\; \ce{\textit{x1} ->[\textcolor{blue}{\textit{(x4)}}] \textit{x2+x3}},\; \ce{\textit{x2+x3} <=>[\textcolor{blue}{\textit{(x1)}}] \textit{x4}},\; \ce{\textit{x4} ->[\textcolor{blue}{\textit{(x3)}}] \textit{x1}},\; \ce{\textit{x4} <=>[\textcolor{blue}{\textit{(x4)}}] \textit{xy}}\}$  and $R^{(2)} = \{\ce{\textit{f1+f2} ->[\textcolor{blue}{\textit{(y3)}}] \textit{y1}},\; \ce{\textit{f1+f2} ->[\textcolor{blue}{\textit{(y1)}}] \textit{y2}}, \; \ce{\textit{y2+y3} <=>[\textcolor{blue}{\textit{(y2)}}] \textit{y3}}\}$.}
\end{figure} 

\section{\label{CRN_RAF}Reflexively Autocatalytic and Food-generated Chemical Reaction Networks}

In this Section we study RAF sets in the context of CRN theory.
Our first step is to represent F-generated sets in terms of CRNs (Section \ref{Food_CRN}).
In particular, we focus our attention on 
zero deficiency CRNs (this restriction is dropped when we introduce catalysis, see Section \ref{Cat_CRN}).
We open the system to the inflow of matter by introducing pseudo reactions that produce species of the food set.
We consider both reversible and irreversible reactions and allow only elementary reactions (i.e.,  reactions involving one or two chemical species as reactants) to occur\footnote{Reactions involving three or more molecules as reactants are rare; they are often introduced as an approximation of sequences of elementary reactions \cite{Gillespie2007}. By excluding non-elementary reactions, the model we propose is suitable for being studied through stochastic simulations \cite{Gillespie1976,Gillespie1977}, without the need to introduce further assumptions.}.

Finally, in Section \ref{Cat_CRN} we introduce the notion of catalysis within the framework of the CRN theory. 
In RAF theory, a catalyst is a species able, with its presence, to facilitate (i.e. accelerate) the proceeding of a reaction \cite{Hordijk2004}.
However, in real systems, catalysed reactions usually involve one or more intermediate complexes and can exhibit various dynamical behaviours \cite{Michaelis1913,Ricard1987}.

We define catalysis within CRNs by introducing an enzymatic mechanism such that a catalysed reactions consumes a catalyst $x$ and a substrate $s$ to form an intermediate complex $e$, which in turn releases product species and the original catalyst $x$:
\begin{equation}
 \ce{\textit{s + x} <=>[\kappa_1][\kappa_{-1}] e <=>[\kappa_2][\kappa_{-2}] \textit{p + x}}
\end{equation}
where $\kappa_{-2} = 0$ for irreversible reactions.
Note that this modeling is the common \textit{Michaelis-Menten scheme} \cite{Michaelis1913}, extensively used to describe enzymatic processes, and that this modelisation has already been used to simulate RAF sets dynamics \cite{Serra2019}.
Furthermore, we assume the validity of the \textit{total quasi steady-state approximation} (tQSSA) \cite{Briggs1925,Tzafriri2003,Tzafriri2004}, that is, the intermediate complex $e$ disappears virtually as fast as it is formed (due to its highly reactivity). In particular, by imposing
\begin{align}
 &\frac{\kappa_{-1}+\kappa_2}{\kappa_1} >> 1,\label{qssa1}\\
 &\frac{\kappa_{-1}+\kappa_2}{\kappa_1} >> \frac{\kappa_2}{\kappa_1},\label{qssa2}\\
 &\frac{\kappa_{-2}}{\kappa_1} << 1,\label{qssa3}
\end{align}
it results \cite{Tzafriri2003,Tzafriri2004}:
\begin{equation}\label{qssa}
 \dot{c}(e) \approx 0. 
\end{equation}
This assumption allows us to provide a relation between the topology of the CRN without catalysis and the corresponding dynamics including enzymatic processes (Appendix~\ref{Apx}).

\subsection{\label{Food_CRN}Food-generated Chemical Reaction Networks}

Let $F$ be a set of \textit{chemostatted} chemical species, that is, a set of species whose concentration is controlled by the environment.
Given a CRN $\mathcal{G} = (\mathcal{S}, \mathcal{N}, \mathcal{R})$, we consider $F \subset S$ to be the food set of $\mathcal{G}$.
As in RAF theory, we say that $\mathcal{G} = (\mathcal{S}, \mathcal{N}, \mathcal{R})$ is a \textit{F-generated CRN} if, for a certain set $F \subset S$ and for each $s \in S$, there exists a sequence of reactions $(\nu'_0-\nu_0),(\nu_1'-\nu_1),\dots,(\nu'_i-\nu_i)$ such that:
\begin{itemize}
 \item $\nu'_i(s) \neq 0$;
 \item $\nu_0(s') \neq 0 \Leftrightarrow s' \in F$;
 \item $\forall s'\in S$ such that $\nu_j(s') \neq0, j=1,\dots,i$ exists $k<j$ such that $\nu'_k(s') \neq 0$. 
\end{itemize}
Let $\mathcal{G} = (\mathcal{S},\mathcal{N},\mathcal{R})$ be a F-generated CRN. 
We say that $\mathcal{G}$ is \textit{minimal} if the following condition holds: 
\begin{equation}\label{minimal}
 \forall (\nu,\mu) \in {\mathcal{N}} \exists s \in \mathcal{S}| \mu \neq \nu \Rightarrow \nu(s) \neq 0, \mu(s) = 0,  
\end{equation}
that is, a F-generated CRN $\mathcal{G}$ is minimal if all its complexes contain at least one species that does not appear in any other complex of the network.
It can be shown that a minimal F-generated CRN is a zero deficiency network.
In fact, let $\mathcal{G}'$ be a minimal F-generated CRN having just one linkage class that connects all its $n$ complexes through exactly $n - 1$ reactions. From Eq.~\ref{minimal} it follows that the reaction vectors $(\nu' - \nu)$ are linear independent. It follows that $d_{\mathcal{G}'} = n - 1$, and it is $\delta_{\mathcal{G}'} = 0$.
Let $\mathcal{G}$ be a CRN obtained from $\mathcal{G}'$ by removing exactly $m$ reactions. Note that $\mathcal{G}$ is again minimal. It results that $l_{\mathcal{G}} = 1+m$, $d_\mathcal{G} = n -1 - m$, and $\delta_{\mathcal{G}} = 0$.
From the arbitrary choice of $m$ and the equivalence of the deficiency for CRNs having the same complexes and the same linkage classes \cite{Feinberg1995}, it follows that each CRN satisfying Eq.~\ref{minimal} is a deficiency zero network.

In order to model the chemostatting of species in $F$, we introduce pseudo reactions for each species $f \in F$ such that:
\begin{equation}
 \ce{\textit{0} <=>[\iota_f][o_f] \textit{f}}
\end{equation}
where $\iota_f$ and $o_f$ are the constant rates for reactions $0 \rightarrow f$ and $f \rightarrow 0$, respectively.
We denote with $\mathcal{G}_F = (\mathcal{S}_F,\mathcal{N}_F,\mathcal{R}_F)$ the CRN such that $\mathcal{S}_F = F$, $\mathcal{N}_F = F \cup 0$ (where $0$ is the zero complex) and $\mathcal{R}_F$ is the set of the pseudo reactions.
Note that for each pair of reaction vectors $(\nu'_i-0),(\nu'_j - 0) \in \mathcal{R}_F$, for $i \neq j$, it is $(\nu'_i-0) \cdot (\nu'_j-0) = 0$, where $\cdot $ is the scalar product.
Thus, it results $d_{\mathcal{G}_F} = |F|$ and $\delta_{\mathcal{G}_F} = 0$, that is, $\mathcal{G}_F$ is a (weakly) reversible zero deficiency CRN.
The DZT states that $\mathcal{G}_F$ admits exactly one steady state. In particular, for each $f \in F$, the positive equilibrium is given by $c(f) = \iota_f / o_f$.
Therefore, $c(f)$ is the concentration that the species $f$ would reach if only only pseudo reactions are included.

Given a CRN $\mathcal{G'} =(\mathcal{S}',\mathcal{N}',\mathcal{R}')$ and a food set $F \subset \mathcal{S}'$, the equivalent \textit{open CRN} $\mathcal{G}$ of $\mathcal{G}'$ is obtained from $\mathcal{G}'$ by adding $\mathcal{G}_F$, such that it is $\mathcal{G} = (\mathcal{S},\mathcal{N},\mathcal{R})$, where $\mathcal{S} = \mathcal{S}'$, $\mathcal{N} = \mathcal{N}' \cup \mathcal{N}_F$ and $\mathcal{R} = \mathcal{R}' \cup \mathcal{R}_F$.
Note that, in general, the addition of $\mathcal{G}_F$ to $\mathcal{G}'$ changes the topology of $\mathcal{G}'$; for instance, $\mathcal{G}_F$ can connect separated linkage classes of $\mathcal{G}'$.
Moreover, the introduction of various chemostatted chemical species affects the dynamical behaviour of the network, both by breaking conservation laws and with the emergence of irreversible cycles \cite{Polettini2014}. 
In this work, we focus our attention on CRNs having no complexes including both food and non-food species, and we do not allow food species to be catalysts. Therefore, for each complexes $\nu \in \mathcal{N}$ and species $s \in \mathcal{S}$ it is:
\begin{align}
 &\nu(f) \neq 0,\nu(s) \neq 0 \Rightarrow s\in F \label{F_cond1}\\
 &(s,\nu \rightarrow \nu')  \in C \Rightarrow s \notin F,\label{F_cond2}
 \end{align}
where $f$ denotes a food species.
Authors are currently investigating the impact of chemostatting on network dynamics if Eq.~\ref{F_cond1} and Eq.~\ref{F_cond2} are not satisfied.

\subsection{\label{Cat_CRN}Addition of catalysis}

\begin{figure*}
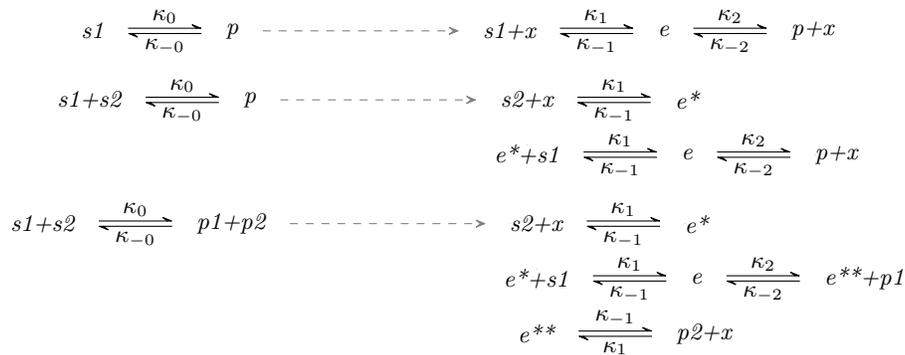

\centering
\schemestart
    \chemfig{\textit{s1}}
    \arrow{<=>[$\kappa_{0}$][$\kappa_{-0}$]}
    \chemfig{\textit{p}}
    \arrow[0,2,dashed,gray]
    \chemfig{\textit{s1+x}}
    \arrow{<=>[$\kappa_{1}$][$\kappa_{-1}$]}
    \chemfig{\textit{e}}
    \arrow{<=>[$\kappa_{2}$][$\kappa_{-2}$]}
    \chemfig{\textit{p+x}}
\schemestop
\vspace{3mm}
\schemestart
    \chemfig{\textit{s1+s2}}
    \arrow{<=>[$\kappa_{0}$][$\kappa_{-0}$]}
    \chemfig{\textit{p}}
    \arrow[0,2,dashed,gray]
    \chemfig{\textit{s2+x}}
    \arrow(s2--){<=>[$\kappa_{1}$][$\kappa_{-1}$]}
    \chemfig{\textit{e*}}
    \arrow(@s2--){0}[270,0.3]
    \chemfig{\textit{e*+s1}}
    \arrow{<=>[$\kappa_{1}$][$\kappa_{-1}$]}
    \chemfig{\textit{e}}
    \arrow{<=>[$\kappa_{2}$][$\kappa_{-2}$]}
    \chemfig{\textit{p+x}}
\schemestop
\vspace{3mm}
\schemestart
    \chemfig{\textit{s1+s2}}
    \arrow{<=>[$\kappa_{0}$][$\kappa_{-0}$]}
    \chemfig{\textit{p1+p2}}    \arrow[0,2,dashed,gray]
    \chemfig{\textit{s2+x}}
    \arrow(s2--){<=>[$\kappa_{1}$][$\kappa_{-1}$]}
    \chemfig{\textit{e*}}
    \arrow(@s2--){0}[270,0.3]
    \chemfig{\textit{e*+s1}}
    \arrow(s1--){<=>[$\kappa_{1}$][$\kappa_{-1}$]}
    \chemfig{\textit{e}}
    \arrow{<=>[$\kappa_{2}$][$\kappa_{-2}$]}
    \chemfig{\textit{e**+p1}}
    \arrow(@s1--){0}[270,0.3]
    \chemfig{\textit{e**}}
    \arrow{<=>[$\kappa_{-1}$][$\kappa_{1}$]}
    \chemfig{\textit{p2+x}}
\schemestop
\caption{Catalysis scheme. The dashed gray arrows associate spontaneous reactions (terms on the left) with the corresponding catalysed reactions (terms on the right).
Catalysis associated with irreversible reactions can be obtained from the first two catalysis of the scheme by setting $\kappa_{-0}$ and $\kappa_{-2}$ equal to zero. Note that for irreversible reactions, $p$ is in turn a complex, i.e., it can be constituted by one or more chemical species.
The associated ODEs are shown in Appendix~\ref{Apx}.}
\label{scheme}
\end{figure*}

In this Section we introduce a procedure to include catalysis within a CRN.
In RAF theory, a catalysis is a pair $(x,r)$ indicating that species $x$ catalyses reaction $r$.
As previously indicated, in our model, we add a catalysis as a substrate-catalyst interaction that produces a \textit{substrate-catalyst complex}. 
In particular, for a monomolecular reaction $s_1 \rightarrow p$ catalysed by a molecule $x$, we assume that the species $s_1$ reacts with the catalyst $x$ to form the substrate-catalyst complex $e$, which can perform the inverse reaction or release the product $p$ and the catalyst $x$.
Similarly, for a bimolecular reaction ${s_1 + s_2 \rightarrow p}$ catalysed by $x$, we assume that one reactant, says $s_2$, is the substrate that reacts with $x$ to produce $e^*$, while the other reactant, $s_2$, reacts with $e^*$ producing $e$, that eventually releases $x$ and products $p$.
Fig.~\ref{scheme} sketched the procedure.

Note that the substrate-catalyst complexes $e,e^*,e^{**}$ (Fig.~\ref{scheme}) are considered as different complexes, that is:
\begin{align}
&(e_i \cdot e_j), (e^*_i \cdot e^*_j), (e^{**}_i \cdot e^{**}_j) \neq 0 \Leftrightarrow i = j\label{e_e}\\
&e_i \cdot e^*_j = e_i \cdot e^{**}_k = e^*_j \cdot e^{**}_k = 0 \; \forall i,j,k.\label{e_e*}
\end{align}
Moreover, it is important to underline that original spontaneous reactions are still present in the network with addition of catalysis.
Given a CRN $\mathcal{G}$ and a set of catalysis $C$, we call $\mathcal{G}_C$ the CRN obtained from $\mathcal{G}$ by adding reactions according to the scheme of Fig.~\ref{scheme} for each catalysis in $C$.
Therefore, $\mathcal{G}_C$ represents the ``real'' mass-action system underlying a CRN to which a set of catalysis is associated.

As in RAF theory, given a F-generated CRN $\mathcal{G} = (\mathcal{S}, \mathcal{N}, \mathcal{R})$ and a set of catalysis $C= \{(s,\nu \rightarrow \nu'): s\in \mathcal{S}, \nu \rightarrow \nu' \in \mathcal{R}\}$, we say that $\mathcal{G}$ is a \textit{RAF CRN} if, for each reaction $\nu \rightarrow \nu' \in \mathcal{R}$, there exists at least one species $x \in \mathcal{S}$ such that $(x,\nu \rightarrow \nu') \in C$; we say that a RAF CRN is a \textit{closed RAF CRN} if, for each reaction $\nu \rightarrow \nu'$ such that all its reactants and at least one catalyst are either part of the set $F$ or are produced by a reaction from $\mathcal{R}$, it $\nu \rightarrow \nu' \in \mathcal{R}$; finally, we say that a closed RAF CRN is a \textit{minimal closed RAF CRN} if it does not contain any other closed RAF CRNs.

\section{\label{motif}Long-term behaviour}

In this Section, we identify motifs in the top logy of a minimal closed RAF CRN $\mathcal{G}$ that allow to predict the dynamics of its associated network $\mathcal{G}_C$ (note that, although not specified, we are considering open CRNs).
In general, adding catalysis increases the deficiency of a CRN \cite{Polettini2015}, and higher deficiency theories are necessary in order to study the steady states of the network \cite{Feinberg1995,Ellison1998}.

However, the topology of the zero deficiency CRN $\mathcal{G}$ provides some information on the dynamics of $\mathcal{G}_C$, at least if conditions (\ref{qssa1}), (\ref{qssa2}) and (\ref{qssa3}) hold (see Appendix~\ref{Apx} for more details).
In fact, the relation among complexes provided by $\mathcal{G}$ does not change in $\mathcal{G}_C$: the terms $\alpha_{\nu \rightarrow \nu'} (\nu - \nu')$ in the sum of Eq.~\ref{ODE} for the two networks, differ only in the values of $\alpha_{\nu \rightarrow \nu'}$, with $\alpha_{\mathcal{G}_C} > \alpha_{\mathcal{G}}$  (Appendix~\ref{Apx}).
This implies, for instance, that if $\mathcal{G}$ is not weakly reversible, thanks to the DZT we can say that the reaction vectors of $\mathcal{G}_C$ are not positively dependent; on the other hand, if $\mathcal{G}$ is weakly reversible, $\mathcal{G}_C$ admits the existence of a positive equilibrium.
Nevertheless, it is worth to underline that $\mathcal{G}$ and $\mathcal{G}_C$ are not equivalent.
For instance, consider the CRN shown in Fig.~\ref{FIG:3a}.
The network without catalysis $\mathcal{G}$ is a weakly reversible zero deficiency network, and it admits exactly one positive equilibrium for each choice of initial condition.
Conversely, by analysing the the associated CRN $\mathcal{G}_C$ through
 the \textit{CRN Toolbox} \cite{Ji2011}, we find that $\mathcal{G}_C$ admits the existence of multiple stationary solutions starting from the same initial conditions, for an appropriate choice of the constant rates $\kappa$.

We use these results in order to predict the dynamics of the mass-action system $\mathcal{G}_C$ associated with a RAF CRN $\mathcal{G}$.
In particular, let $\mathcal{G}$ be a minimal closed RAF CRN, and let $\mathcal{G}_{-}$ be the CRN obtained from $\mathcal{G}$ by removing all the reactions $\nu_0 \rightarrow \nu'$, where the complex $\nu_0$ contains (only) food species.
We define:
\begin{enumerate}
 \item \textit{Fully connected} (FC) motif: $\mathcal{G}$ is a FC motif if it is a weakly reversible CRN;
 \item \textit{Non connected} (NC) motif: $\mathcal{G}$ is a NC motif if both $\mathcal{G}$ and $\mathcal{G}_{-}$ are not weakly reversible CRNs;
 \item \textit{Core connected} (CC) motif: $\mathcal{G}$ is a CC motif if it is not a weakly reversible CRN and $\mathcal
 {G}_{-}$ is a weakly reversible CRN.\label{cIII} 
\end{enumerate}

Fig.~\ref{FIG:3} shows examples of the three motifs.
They exhibit dynamics with different characteristics (Fig.~\ref{FIG:4a}).
In particular, we focus our attention on three dynamical properties of the concentrations of species that constitute a motif:
\begin{enumerate}
 \item \textit{homeostasis}: the ratio among concentrations of all the species is constant over time \label{ho};
 \item \textit{continuous growth}: the concentrations of all the species increase in time\label{cg};
 \item \textit{self-conservation}: if the inflow of matter towards the CRN is interrupted, the concentrations of all the species do not decrease in time \label{sc}.
\end{enumerate}
Note that these are desired features for a system at the basis of the origin of life.
For instance, homeostasis is considered to be a fundamental property connected with self-maintenance of an organism; a growth of species concentration is necessary for self-reproduction, and self-conservation prevents the decay of the system in case of insufficient resources \cite{Luisi2016,Varela2000}.

We argue that FC motif exhibits homeostasis, NC motif exhibits none of the properties and CC motif exhibits continuous growth and self-conservation. It is important to underline that we are referring to long-term dynamics of the networks and not to their transient behaviour.

In fact, FC motif admits the existence of a positive equilibrium for the concentration of all its chemical species, meaning that, for $t \rightarrow \infty$, the ratio $c(s)/c(s')$ is constant for each pair $s,s' \in \mathcal{S}$ (Fig.~\ref{FIG:4a}).
However, if the inflow rate $\iota_f$ is set equal to zero for each $f \in F$, the resulting network is no more weakly-reversible.
In particular, there is a terminal strong-linkage class (i.e., the zero complex) that is an \textit{absorbing linkage class}, meaning that, once the flow of matter falls in it, it can no longer leave it.
Therefore, the matter consumed by reactions involving the species of the CRN is not replaced, the species concentrations decrease and the network is not self-conservative (Fig.~\ref{FIG:4a}).

On the other hand, NC motif contains an absorbing linkage class involving a strict subset $\mathcal{S}'$ of $\mathcal{S}$, both in case of $\iota_f >0$ and $\iota_f = 0$.
As a result, species in $\mathcal{S}'$ increase their concentrations at the expense of other species $s \notin \mathcal{S}'$.
Thus, even if a subset of $\mathcal{S}$ can experience continuous growth, the whole NC motif does not satisfy any of the properties listed above (Fig.~\ref{FIG:4a}).

By definition, instead, all the non-food species of a CC motif belong to a weakly reversible (sub)network that is absorbing for the inflowing matter coming from the environment.
This implies that, if $\iota_f > 0$, each non-food species is reached by a positive flow of matter (continuous growth); on the other hand, if $\iota_f = 0$, the matter is redistributed among non-food species and does not escape from the CRN (self-conservation).
Note that a CC motif could exhibit homeostasis, depending on the fine details of its topology.

\begin{figure*}
\begin{subfigure}[b]{0.3\textwidth}
\schemestart
    \chemfig{\textit{f1}}
    \arrow{<=>}
    \chemfig{\textit{0}}
    \arrow{<=>}
    \chemfig{\textit{f2}}
\schemestop
\\ \quad \\ 
\schemestart
    \chemfig{\textit{f1+f2}}
    \arrow{->[\textcolor{blue}{\textit{(x2)}}]}
    \chemfig{\textit{x1}}
    \arrow{->[*{0}\textcolor{blue}{\textit{(x4)}}]}[270]
    \chemfig{\textit{x2+x3}}
    \arrow{<=>[\textcolor{blue}{\textit{(x1)}}]}[180]
    \chemfig{\textit{x4}}
    \arrow{->[*{0}\textcolor{blue}{\textit{(x3)}}]}[90]
\schemestop
\centering\caption{}\label{FIG:3a}
\end{subfigure}
\hfill
\begin{subfigure}[b]{0.3\textwidth}
\schemestart
    \chemfig{\textit{f1}}
    \arrow{<=>}
    \chemfig{\textit{0}}
    \arrow{<=>} 
    \chemfig{\textit{f2}}
\schemestop
\\ \quad \\ 
\schemestart
    \chemfig{\textit{f1+f2}}
    \arrow{->[\textcolor{blue}{\textit{(x2)}}]}
    \chemfig{\textit{x1}}
    \arrow{->[*{0}\textcolor{blue}{\textit{(x4)}}]}[270]
    \chemfig{\textit{x2+x3}}
    \arrow{<=>[\textcolor{blue}{\textit{(x1)}}]}[180]
    \chemfig{\textit{x4}}
\schemestop
\centering\caption{}\label{FIG:3b}
\end{subfigure}%
\hfill
\begin{subfigure}[b]{0.3\textwidth}
\schemestart
    \chemfig{\textit{f1}}
    \arrow{<=>}
    \chemfig{\textit{0}}
    \arrow{<=>} 
    \chemfig{\textit{f2}}
\schemestop
\\ \quad \\ 
\schemestart
    \chemfig{\textit{f1+f2}}
    \arrow{->[\textcolor{blue}{\textit{(x2)}}]}
    \chemfig{\textit{x1}}
    \arrow{->[*{0}\textcolor{blue}{\textit{(x4)}}]}[270]
    \chemfig{\textit{x2+x3}}
    \arrow{<=>[\textcolor{blue}{\textit{(x1)}}]}[180]
    \chemfig{\textit{x4}}
    \arrow{->[*{0.0}\textcolor{blue}{\textit{(x3)}}]}[40,1.6]
\schemestop
\centering\caption{}\label{FIG:3c}
\end{subfigure}%
\hfill
\\\quad\\
\begin{subfigure}[b]{0.3\textwidth}
\schemestart
    \chemfig{\textit{f1}}
    \arrow{<=>}
    \chemfig{\textit{0}}
    \arrow{<=>} 
    \chemfig{\textit{f2}}
\schemestop
\\ \quad \\ 
\schemestart
    \chemfig{\textit{y1}}
    \arrow{<-[\textcolor{blue}{\textit{(y3)}}]}
    \chemfig{\textit{f1+f2}}
    \arrow{->[\textcolor{blue}{\textit{(y1)}}]}
    \chemfig{\textit{y2}}
\schemestop
\\ \quad \\ 
\schemestart
    \chemfig{\textit{y1+y2}}
    \arrow{<=>[\textcolor{blue}{\textit{(y2)}}]}[0]
    \chemfig{\textit{y3}}
\schemestop
\centering\caption{}\label{FIG:3d}
\end{subfigure}%
\hfill
\begin{subfigure}[b]{0.3\textwidth}
\schemestart
    \chemfig{\textit{f1}}
    \arrow{<=>}
    \chemfig{\textit{0}}
    \arrow{<=>} 
    \chemfig{\textit{f2}}
\schemestop
\\ \quad \\
\schemestart
    \chemfig{\textit{y1}}
    \arrow{<-[\textcolor{blue}{\textit{(y3)}}]}
    \chemfig{\textit{f1+f2}}
    \arrow{->[\textcolor{blue}{\textit{(y1)}}]}
    \chemfig{\textit{y2}}
\schemestop
\\ \quad \\ 
\schemestart
    \chemfig{\textit{y1+y2}}
    \arrow{<=>[\textcolor{blue}{\textit{(y2)}}]}[0]
    \chemfig{\textit{y3}}
    \arrow{->[\textcolor{blue}{\textit{(y3)}}]}
    \chemfig{\textit{y4}}
\schemestop
\centering\caption{}\label{FIG:3e}
\end{subfigure}%
\hfill
\begin{subfigure}[b]{0.3\textwidth}
\schemestart
    \chemfig{\textit{0}}
    \arrow(0--f1){<=>}[170]
    \chemfig{\textit{f1}}
    \arrow(@0--f2){<=>}[190] 
    \chemfig{\textit{f2}}
    \arrow(@0--f3){<=>}[10]
    \chemfig{\textit{f3}}
    \arrow(@0--f4){<=>}[350] 
    \chemfig{\textit{f4}}
\schemestop
\\ \quad \\ \quad \\
\schemestart
    \chemfig{\textit{y1}}
    \arrow(y1--){<-[\textcolor{blue}{\textit{(y3)}}]}[0,0.6]
    \chemfig{\textit{f1+f2}}
    \arrow{->[\textcolor{blue}{\textit{(y1)}}]}[0,0.6]
    \chemfig{\textit{y2}}
    \arrow(@y1--){<-[\textcolor{blue}{\textit{(y3)}}]}[180,0.6]
    \chemfig{\textit{f3+f4}}
\schemestop
\\ \quad \\ 
\schemestart
    \chemfig{\textit{y1+y2}}
    \arrow{<=>[\textcolor{blue}{\textit{(y2)}}]}[0]
    \chemfig{\textit{y3}}
\schemestop
\centering\caption{}\label{FIG:3f}
\end{subfigure}%
\caption{Examples of motifs. Complexes are represented by black letters (the letter $f$ indicates food species), reactions by arrows.
Blue letters in brackets indicate catalysts.
(a) FC motif: the CRN is weakly reversible. 
(b,e) NC motif: the CRN is not weakly reversible, and there is not a weakly reversible (sub)network that includes all the non-food species. 
(c,d,f) CC motif: the CRN is not weakly reversible, and there is a weakly reversible (sub)network that includes all the non-food species.
Note that all the motifs are closed RAF CRNs.} 
\label{FIG:3}
\end{figure*}

\begin{figure*}
\centering
\begin{subfigure}[b]{0.4\textwidth}
		\includegraphics[scale=0.5]{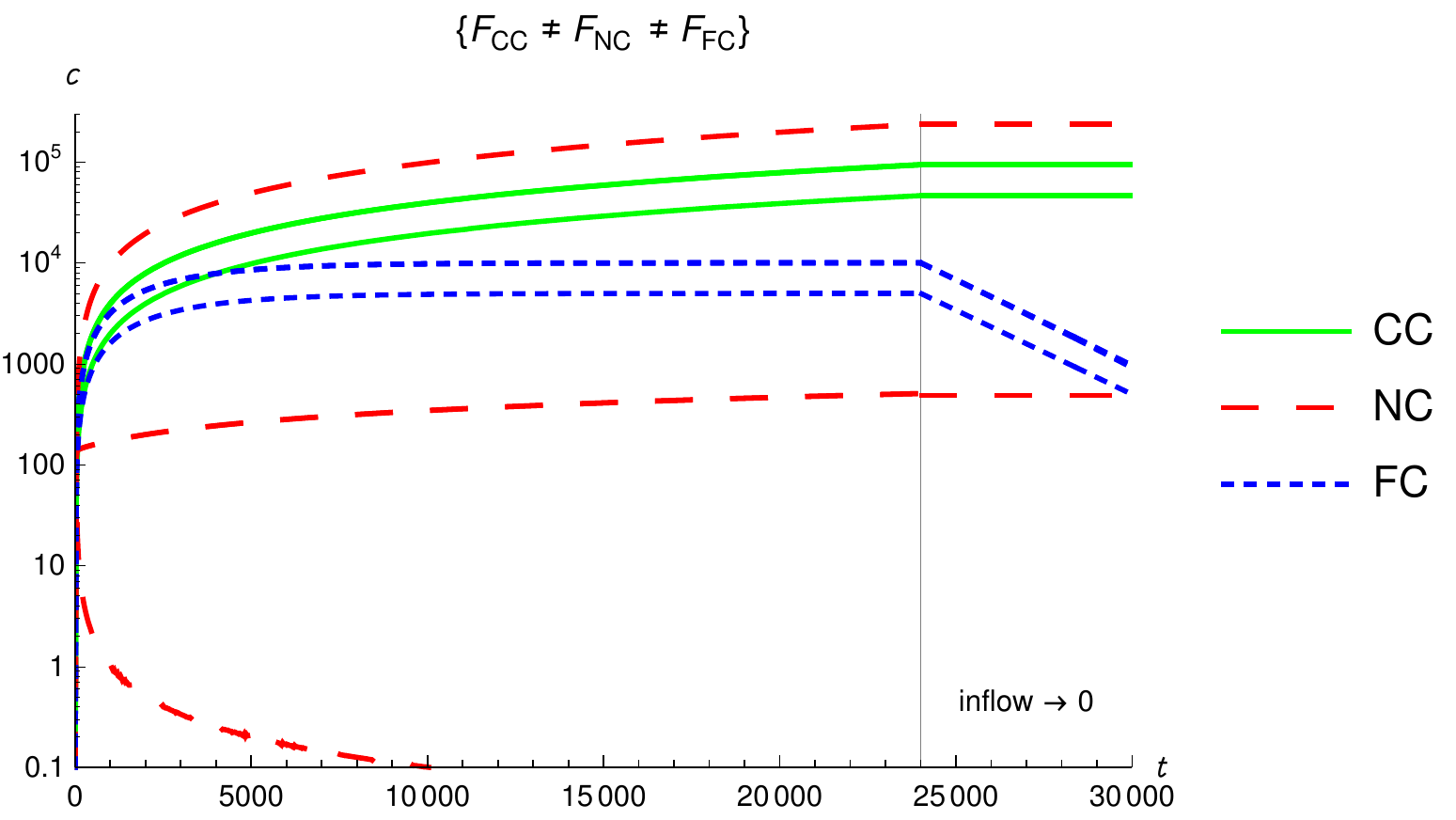}
		\caption{}\label{FIG:4a}
\end{subfigure}
		\hfill
\begin{subfigure}[b]{0.4\textwidth}
		\includegraphics[scale=0.5]{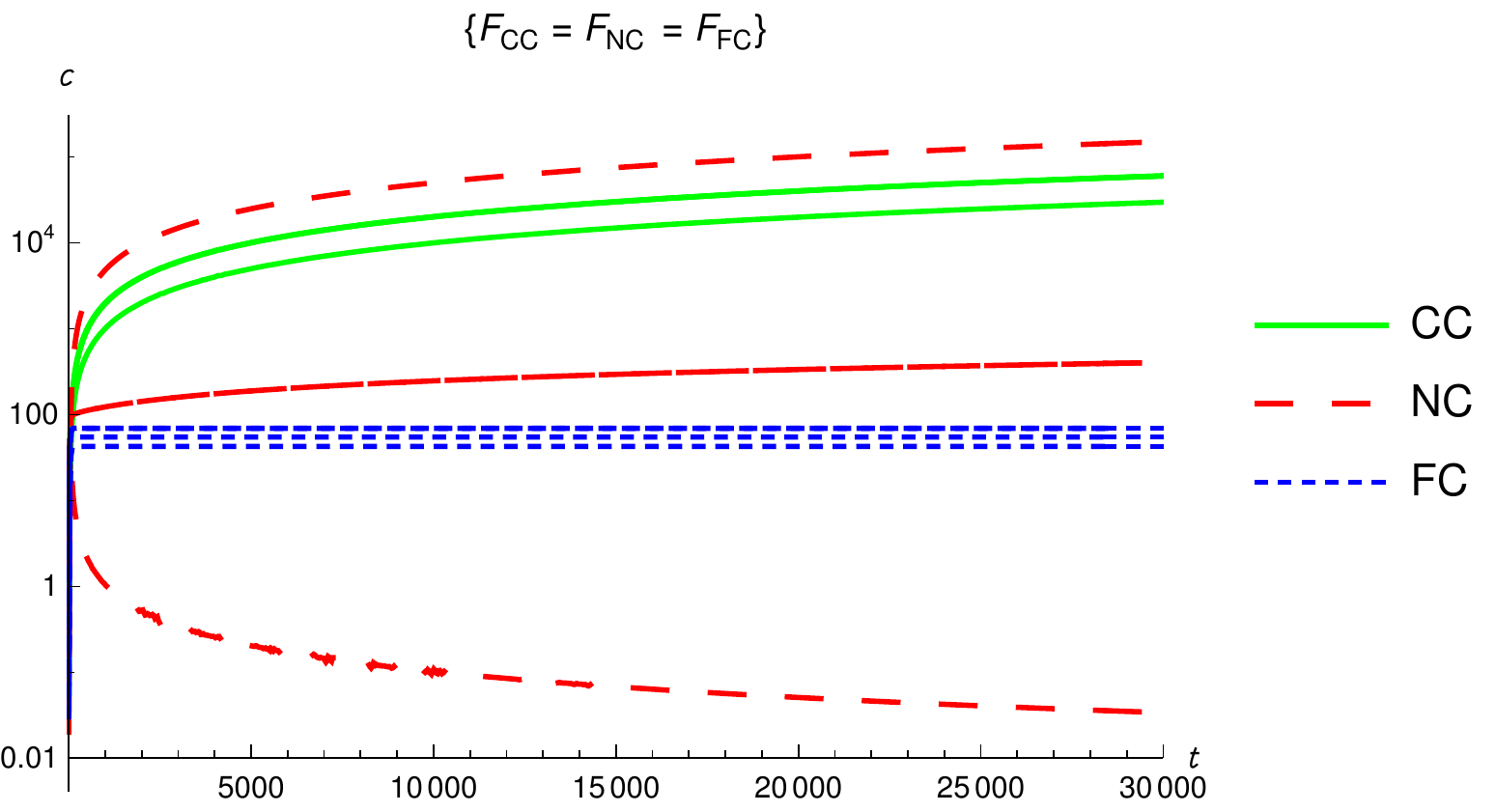}
		\caption{}\label{FIG:4b}
\end{subfigure}
\hfill
\begin{subfigure}[b]{0.4\textwidth}
		\includegraphics[scale=0.5]{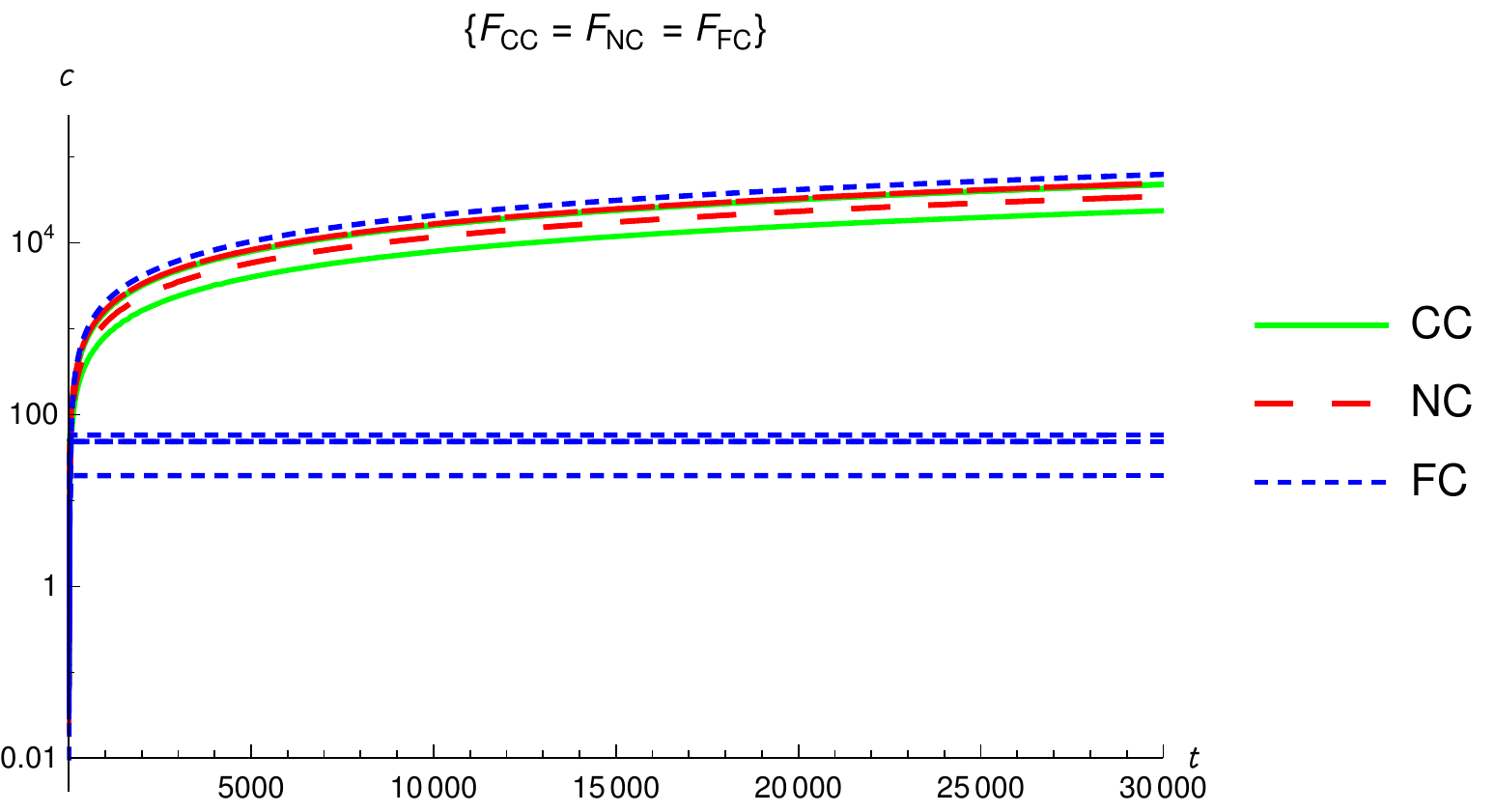}
		\caption{}\label{FIG:4c}
\end{subfigure}
	\caption{Dynamics of motifs. On the $x$-axis: time steps $t$.
	On the $y$-axis: species concentration $c$.
	(a) Dynamics of the three isolated motifs introduced in Fig.~\ref{FIG:3a}, Fig.~\ref{FIG:3b} and Fig.~\ref{FIG:3c}. For $t > 24000$ the inflow rate $\iota_f$ is set equal to $0$ for each pseudo reaction $\ce{\textit{0}  ->[\iota_f} \textit{f}$.
	(b) Dynamics of the three motifs introduced in Fig.~\ref{FIG:3a}, Fig.~\ref{FIG:3b} and Fig.~\ref{FIG:3c}, that draw from the same food set $F = \{f1, f2\}$.
	(c) Resulting dynamics after mutations that change the FC motif of Fig.~\ref{FIG:3a} in a NC motif, and the NC motif of Fig.~\ref{FIG:3b} in a CC motif.
	The plotted curves are obtained by numerically solving the ODEs associated with the CRNs (Eq.~\ref{ODE}), setting $\iota_{f} = 10$, $o_f = 10^{-2}$, $\kappa_0 = \kappa_{-0} = 10^{-3}$, $\kappa_{1} = 10$, $\kappa_{-1} = 10^7$, $\kappa_{2} = 10^4$, $\kappa_{-2} = 10^{-1}$, and an initial concentration of $c_0 = 10^{-2}$ for all the species.}
	\label{FIG:4}
\end{figure*}

\subsection{An illustrative example of evolvability of RAF sets}

In this Section we present illustrative examples on how the detected motifs (Section~\ref{motif}) impact the long-term dynamics of a (larger) RAF CRN. 
We start our study by simulating the dynamics of a RAF CRN constituted by three different closed RAF CRNs that share the same food set $F$.
The closed RAF CRNs correspond to the different motifs we have introduced in Fig.~\ref{FIG:3}.

Fig.~\ref{FIG:4b} shows the dynamics of this particular network: as expected, the resulting dynamics is substantially equivalent to that shown by the non composed motifs (Fig.~\ref{FIG:4a}, $\iota_f > 0$).
In fact, letting the motifs share the food set does not affect their structures, nor does significantly change the ODEs describing the dynamics (apart from a change in the equilibrium constants of FC motif).
Therefore, the overall dynamics of the network is given by the union of the dynamics of the single motifs.

However, introducing new reactions among species can have a major role in the resulting dynamics of the network.
For instance, adding the reaction $\ce{\textit{x4} ->[\textcolor{blue}{\textit{(x1)}}] \textit{x5} }$ to the FC motif of Fig.~\ref{FIG:3a} changes it into a NC motif.
At the same time, adding the reaction $\ce{\textit{x4} ->[\textcolor{blue}{\textit{(x4)}}] \textit{x1}}$ to the NC motif of Fig.~\ref{FIG:3b} changes it into a CC motif.
The dynamics of the overall network, in fact, is the result of the evolution of a FC motif and two CC motifs (Fig.~\ref{FIG:4c}).
This is an example of a general case: it is always possible to pass from one motif to another by adding (or removing) one or more reactions.

\begin{figure*}
\centering
\begin{subfigure}[b]{0.4\textwidth}
		\includegraphics[scale=0.5]{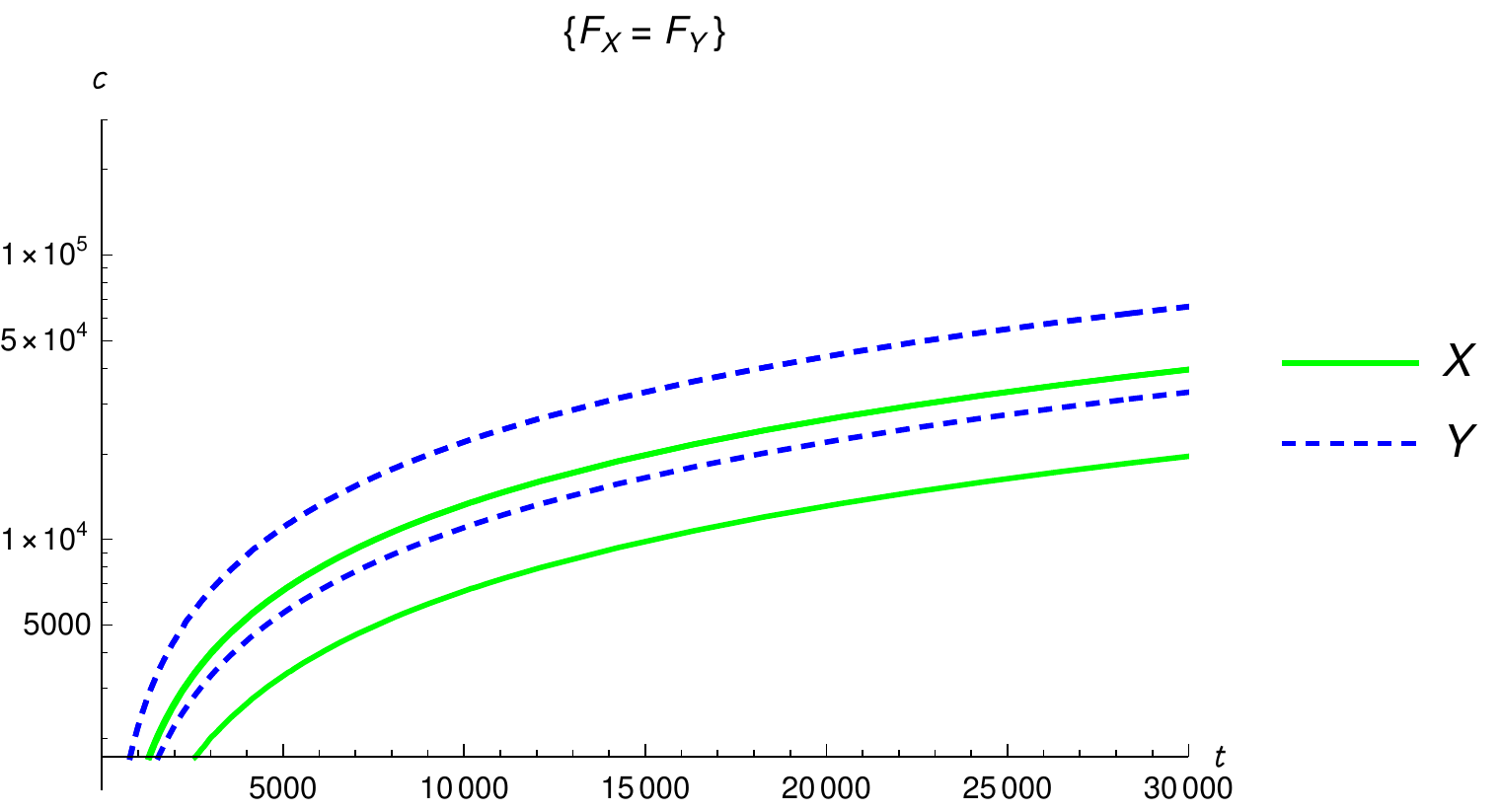}
		\caption{}\label{FIG:5a}
\end{subfigure}
		\hfill
\begin{subfigure}[b]{0.4\textwidth}
		\includegraphics[scale=0.5]{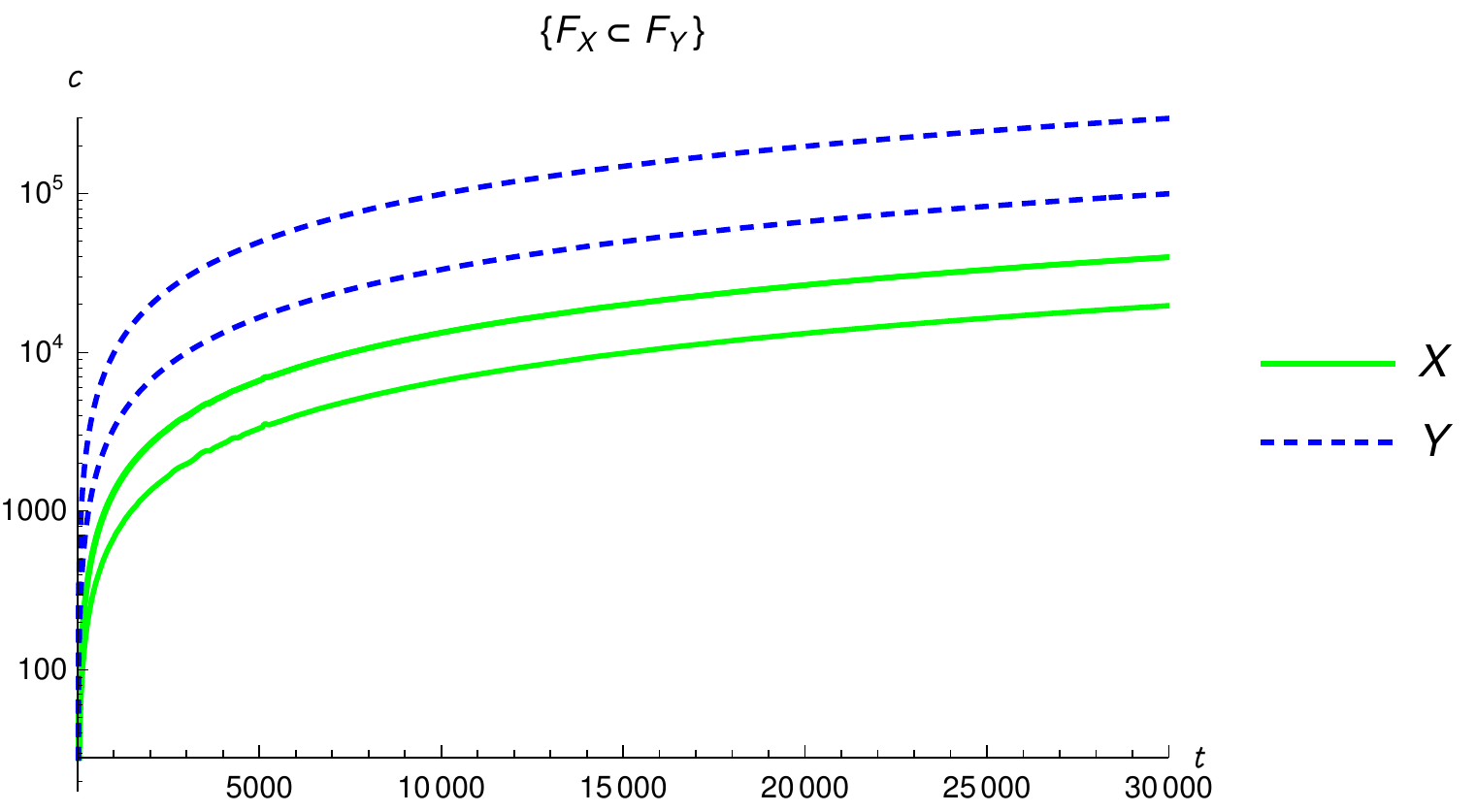}
		\caption{}\label{FIG:5b}
\end{subfigure}
		\hfill
\begin{subfigure}[b]{0.4\textwidth}
		\includegraphics[scale=0.5]{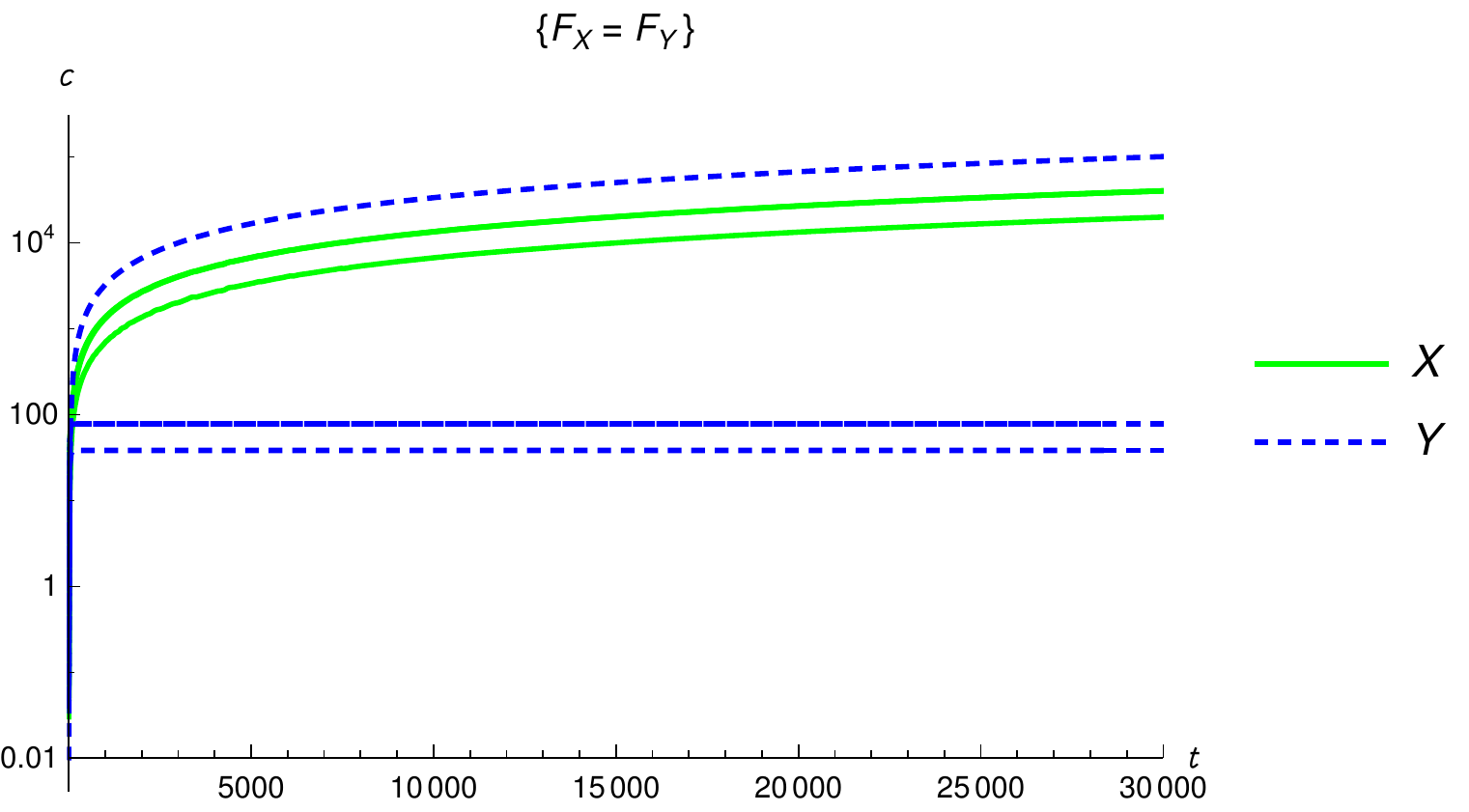}
        \caption{}\label{FIG:5c}
\end{subfigure}
		\hfill
\begin{subfigure}[b]{0.4\textwidth}
		\includegraphics[scale=0.5]{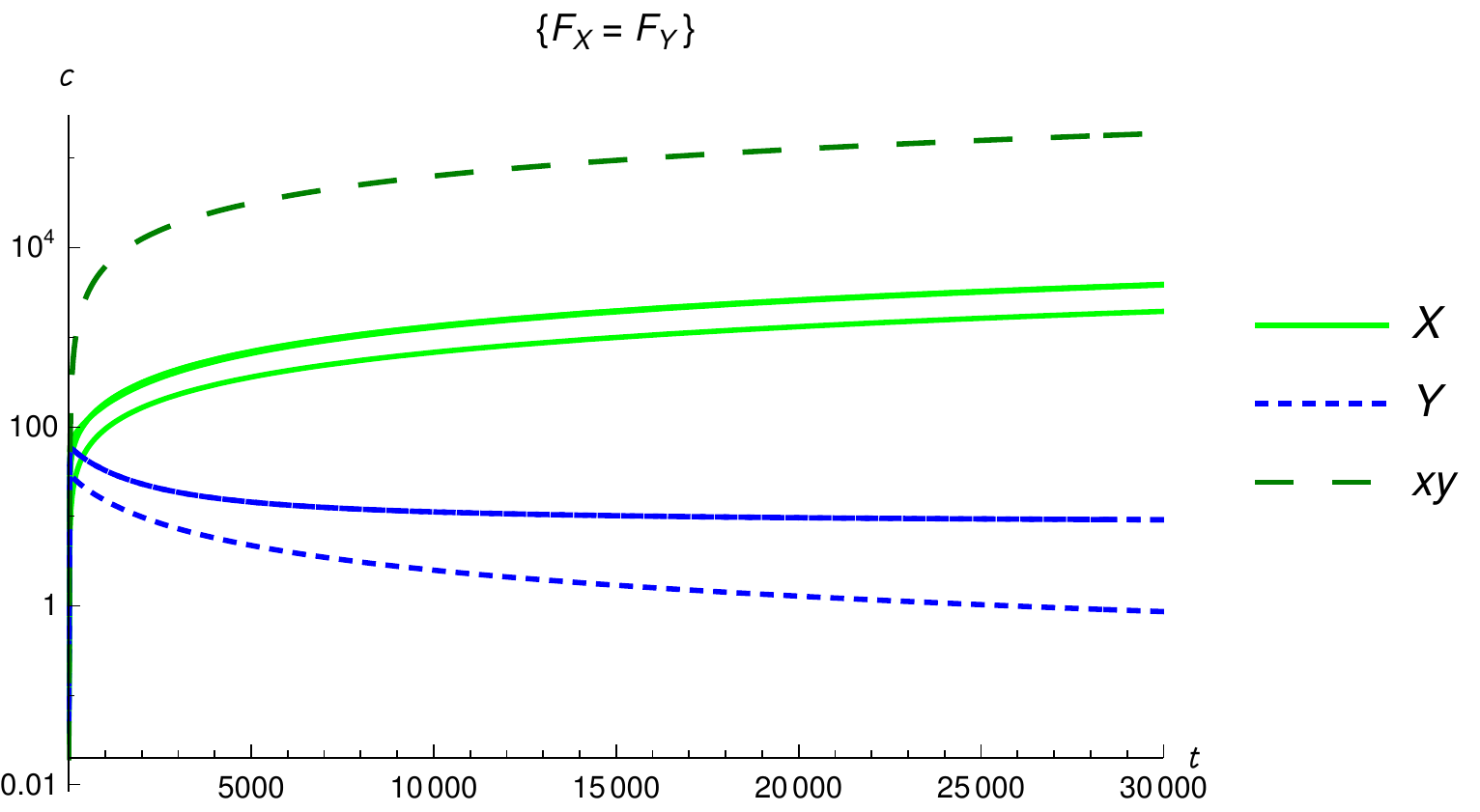}
		\caption{}\label{FIG:5d}
\end{subfigure}
	\caption{Evolution of a RAF CRN induced by mutations. On the $x$-axis: time steps $t$.
	On the $y$-axis: species concentration $c$.
	(a) Dynamics of a RAF CRN constituted by the CC motif of Fig.~\ref{FIG:3c} ($X$) and the CC motif of Fig.~\ref{FIG:3d} ($Y$), that draw from the same food set $F = \{f1,f2\}$. 
	(b) A favorable mutation allows $Y$ to draws resources from the expanded set $\{f1, f2, f3, f4\}$ (Fig.~\ref{FIG:3f}), resulting in a higher self-production rate.
	(c) A unfavorable mutation changes $Y$ in a NC motif (Fig.~\ref{FIG:3e}), preventing from its continuous growth.
	(d) A mutation allows $X$ to steal matter from $Y$, causing the dynamical disappearance of $Y$ (the CRN associated with this dynamics is shown in Fig.~\ref{FIG:1}).
	The plotted curves are obtained by numerically solving the ODEs associated with the CRNs (Eq.~\ref{ODE}), setting $\iota_{f} = 10$, $o_f = 10^{-2}$, $\kappa_0 = \kappa_{-0} = 10^{-3}$, $\kappa_{1} = 10$, $\kappa_{-1} = 10^7$, $\kappa_{2} = 10^4$, $\kappa_{-2} = 10^{-1}$, and an initial concentration of $c_0 = 10^{-2}$ for all the species.}
	\label{FIG:5}
\end{figure*}

This last point has important consequences for the evolvability of RAF sets.
In fact, we can interpret the characteristic dynamical behaviour of a motif as its phenotype, and the (random) addition of a new reaction as a mutation.
In this framework, the examples of Fig.~\ref{FIG:4b} and Fig.~\ref{FIG:4c} show that closed RAF CRNs can acquire (or lose) dynamical properties that can be, potentially, picked by natural selection.
As further evidence to support this observation, in Fig.~\ref{FIG:5} we show different frames of an hypothetical evolution of a RAF CRN in which mutations occur.
In particular, the starting network is formed by two closed RAF CRNs that share the same food set $F = \{f1, f2\}$.
The closed RAF CRNs are indicated with letters $X$ (CC motif shown in Fig.~\ref{FIG:3c}) and $Y$ (CC motif shown in Fig.~\ref{FIG:3d}), respectively.
Initially, both the closed RAF CRNs show a continuous growth of species concentrations, with similar growing rates (Fig.~\ref{FIG:5a}).
However, the addition of a new reaction in the network allow $Y$ to drawn resources from a supplementary source compared to $X$, represented by the food elements $f3$ and $f4$ (Fig.~\ref{FIG:3f}).
Consequently, this favorable mutation provides an advantage for $Y$, which is able to grow faster than $Y$ (Fig.~\ref{FIG:5b}).
At the same time, as previously observed, an unfavorable mutation (Fig.~\ref{FIG:3e}) can decrease the efficiency of a network's self-reproduction (Fig.~\ref{FIG:5c}).
Finally, note that a mutation introducing a connection between $X$ and $Y$ can have major effects on their dynamics.
In particular, consider the RAF CRN shown in Fig.~\ref{FIG:1}: it can be built starting from the networks of Fig.~\ref{FIG:3c} and Fig.~\ref{FIG:3d}, by adding the production of the species $xy$.
This latter expansion of the network can be interpreted, for instance, as the result of a mutation that allows species $x4$ to produce species $xy$ in a reversible way, and species $x1$ to catalyse the irreversible production of species $xy$ starting from species $y3$.
Therefore, thanks to a favorable mutation, one motif can steal resources from the other.
The dynamics of the resulting RAF CRN is shown in Fig.~\ref{FIG:5d}: due to the mutation, the closed RAF CRN $X'$ (the symbol $X'$ indicates the network $X$ to which the reaction $\ce{\textit{x4} ->[\textcolor{blue}{\textit{(x4)}}] \textit{xy}}$ is added) is an absorbing linkage class in which the flow of matter of the net $Y$ falls; consequently, network $X$ still shows a continuous growth in species concentrations, while $Y$ dynamically disappears.

In this Section, we have shown how the presence of motifs FC, NC and CC within the structure of a RAF CRN affects the dynamics of the system.
Moreover, the examples we have introduced suggest that closed RAF CRNs could be the elementary units that experience adaptive evolution.
Regarding this issue, it is important to underline two points.
First, note that, in this work, we are dealing with a deterministic model.
This approximation does not allow to directly simulate neither the spontaneous appearance of novelty, nor the probabilistic effects, that are characteristics of systems with few elements.
However, recent numerical simulations have shown how these stochastic processes can determine both a dependence on different initial conditions (for instance, deriving from an inhomogeneous distribution of resources, or from random events, such as the division of a protocell), and an actual competition among closed RAF CRNs \cite{Hordijk2018b,Serra2019, Ravoni2020}.
Therefore, the addition of stochastic elements to the RAF CRN model introduced in this work should reproduce the scenario proposed in \cite{Vasas2012,Hordijk2018b,Ravoni2020} for the evolvability of a population of compartmentalised RAF sets.

Moreover, it is very important to note that our results show that a RAF CRN does not always fully emerges in the long-term dynamics, and that an actual competition can therefore occur also among subsets of the same RAF network.

We are currently investigating the impact of some of the assumptions made in this work, such as the tQSSA or the limited role of the food set in the network reactions, and will discuss all these aspects in an upcoming work.

\section{\label{conclusions}Conclusions}

In this work, we study long-term behaviours of RAF sets, particularly focusing on those systems that are considered to be the building blocks from which life emerged.

For this purpose, we represent RAF sets in terms of CRNs, and we use CRN theory to predict the dynamics of the networks starting from their topology.
Our main result is the identification of additional topological conditions compared to the one introduced by the RAF theory, which allow us to predict the dynamics of the system and to identify networks with interesting dynamical properties.
To our knowledge, this is the first time that the actual dynamics of RAF sets have been successfully connected with their topology.

Furthermore, starting from their structure, we display that not all the subsets of a RAF network eventually  dynamically emerge, contrary to what has been observed in previous works, where too simple CRNs were considered \cite{Hordijk2018b,Ravoni2020}.
These results show that the increase of complexity of a RAF set significantly affects the emergent dynamics, supporting the potential evolvability of RAF sets and their importance in the transition from inanimate to living matter.

Thermodynamic aspects of these systems are going to be investigated in a forthcoming paper, allowing us to formalise a consistent theory of Autocatalytic networks.

\appendix
 \section{\label{Apx}Appendix}
 
 Let $\mathcal{G} = (\mathcal{S},\mathcal{N},\mathcal{R})$ be a CRN, $\mathcal{R^+}$ and $\mathcal{R^-}$ be the sets of reversible and irreversible reactions (such that $\mathcal{R} = \mathcal{R^+} \cup \mathcal{R^-}$), and  $i,j$ denote reactions $\nu \rightarrow \nu' \in \mathcal{R}$.
 The function $f(c,\kappa)$ for $\mathcal{G}$ can be written as:
 \begin{equation}
  f_{\mathcal{G}}(c,\kappa) = \sum_{i \in \mathcal{R^+}}\kappa_{i} c^{\nu_{i}}(\nu'_{i} -\nu_{i}) + \sum_{j \in \mathcal{R^-}}\kappa_{j} c^{\nu_{j}}(\nu'_{j} -\nu_{j}).
 \end{equation}
 Similarly, for $\mathcal{G}_c$ it is:
 \begin{align}
 f_{\mathcal{G}_C}(c,\kappa) =  &\sum_{i \in \mathcal{R^+}}[\kappa_{i} c^{\nu_{i}}(\nu'_{i} -\nu_{i}) + P_i]+\\+&\sum_{j \in \mathcal{R^-}}[\kappa_{j} c^{\nu_{j}}(\nu'_{j} -\nu_{j}) + I_j],\notag
 \end{align}\label{equilibrium_cond}
 where $P_i$ and $I_j$ denote the additional terms introduced by adding catalysis according to the scheme of Fig.~\ref{scheme} for reversible and irreversible reactions, respectively.
 For each pair $(i,i')$ of reversible reactions such that $i: \nu \rightarrow \nu'$ and $i': \nu'\rightarrow \nu$, the term $P_{(i,i')}$ can be written as:
 \begin{align}
  P_{(i,i')} = & |s2| [\kappa_1 c(s2) c(x)-\kappa_{-1} c(e^*) ] \label{P}\\ 
  \times & (e^* - s2 -x ) + \{\kappa_1 c^{(\nu - s2)} [ |s2| c(e^*) \notag \\
  + & ( 1 - |s2| ) c(x) ]-\kappa_{-1} c(e)\}[e-\nu  \notag \\
  + & s2 - |s2| e^* - ( 1-|s2| ) x] \notag \\
  + & \{\kappa_2 c(e) - \kappa_{-2} c^{ (\nu'-p2)} [ |p2|c(e^{**} ) \notag \\
  + & (1-|p2|)c(x)]\}[\nu'-p2+|p2|e^{**} \notag \\
  + & (1-|p2|)x - e] + |p2|[\kappa_1 c(p2) c(x) \notag \\
  - & \kappa_{-1} c(e^{**})] (e^{**}-p2-x). \notag
   \end{align}
 Here, $s2$ and $p2$ are the vectors having all components equal to zero, except one (corresponding to species $s2$ and $p2$, respectively).
 Note that due to the above definition, $P_{(i,i')}$ describes the terms to be added for all the catalysis shown in Fig.~\ref{scheme}, depending on whether the relations $s2 = 0$ and $p2 = 0$ are valid or not, where $0$ is the zero complex.
 
 Similarly, for each irreversible reaction $j: \nu \rightarrow \nu'$ the term $I_j$ can be written as:
 \begin{align}
  I_{(j)} = & |s2| [\kappa_1 c(s2) c(x)-\kappa_{-1} c(e^*) ] \label{I}\\ 
  \times & (e^* - s2 -x ) + \{\kappa_1 c^{(\nu - s2)} [ |s2| c(e^*) \notag \\
  + & ( 1 - |s2| ) c(x) ]-\kappa_{-1} c(e)\}[e-\nu  \notag \\
  + & s2 - |s2| e^* - ( 1-|s2| ) x] \notag \\
  + & [\kappa_2 c(e)](\nu'+ x - e). \notag
   \end{align}
 Note that $I_j$ can be obtained from $P_{(j,j')}$ assuming $\kappa_{-2} = 0$ and $p2 = 0$.
 From Eq.~\ref{P} it results:
 \begin{align}
  \dot{c}(e) = & \{\kappa_1 c^{(\nu-s2)} [|s2|c(e^*)+(1-|s2|)c(x)] \\ 
  - & \kappa_{-1}c(e)+\kappa_2 c(e)-\kappa_{-2}c^{(\nu'-p2)}[|p2|c(e^{**}) \notag \\
  + & (1-|p2|)c(x)]\}, \notag \\
  \dot{c}(e^*) = & |s2|[\kappa_1 c(s2)c(x)-\kappa_{-1} c(e^*) \\
  - & \kappa_1 c^{(\nu-s2)}c(e^*) +\kappa_{-1} c(e)], \notag \\
  \dot{c}(e^{**}) = & |p2|[\kappa_1 c(p2)c(x)-\kappa_{-1}c(e^{**}) \\
  - & \kappa_{-2}c^{(\nu'-p2)}c(e^{**})+\kappa_2 c(e)]. \notag
 \end{align}
 By imposing $\dot{e}=0$, $\dot{e}^*=0$ and $\dot{e}^{**}=0$ (tQSSA), it follows:
 \begin{align}
  \kappa_2 c(e)-\kappa_{-2}c^{(\nu'-p2)}c(e^{**}) = &\kappa_1 c^{(\nu-s2)}c(e^*) \label{equivalences}\\
  - & \kappa_{-1}c(e). \notag 
 \end{align}
 Again, equivalent expressions for irreversible reactions are obtained assuming $\kappa_{-2} = 0$, $p2 = 0$ and (or) $s2 = 0$.
 Note that the equivalences of Eq.~\ref{equivalences} imply that, if for $f_{\mathcal{G}}$ it is:
 \begin{equation}
 f_{\mathcal{G}} = \sum_{\nu \rightarrow \nu'} \alpha_{\nu \rightarrow \nu'} (\nu'-\nu),
 \end{equation}
 then $f_{\mathcal{G}_C}$ can be written as:
 \begin{equation}
  f_{\mathcal{G}_C} = \sum_{\nu \rightarrow \nu'} \alpha'_{\nu \rightarrow \nu'} (\nu'-\nu),
 \end{equation}
 where:
 \begin{equation}
  \alpha'_{\nu \rightarrow \nu'} = \alpha_{\nu \rightarrow \nu'} + \Delta_{\nu \rightarrow \nu'},
 \end{equation}
 with $\Delta_{\nu \rightarrow \nu'} \geq 0$.
 For instance, if $\kappa_{-2} \neq 0$, $p2 \neq 0$ and $s2 \neq 0$ it is:
 \begin{equation}
  \Delta_{\nu \rightarrow \nu'} = \kappa_2 c(e)
 \end{equation}
 for a forward reaction (that is, a reaction with a ``spontaneous'' rate constant $\kappa_0$) and:
 \begin{equation}
  \Delta_{\nu \rightarrow \nu'} = \kappa_{-2}c^{(\nu'-p2)}c(e^{**})
 \end{equation}
 for a reverse reaction (that is, a reaction with a ``spontaneous'' rate constant $\kappa_{-0}$).
 This implies, in particular, that $\mathcal{G}_C$ admits a positive equilibrium only if $\mathcal{G}$ is weakly reversible. An equivalent result holds if $\kappa_{-2} = 0$, $p2 = 0$ and (or) $s2 = 0$.

\end{document}